\documentclass{ws-rv9x6}
\usepackage{subfigure}   
\usepackage{ws-rv-thm}   
\usepackage{ws-rv-van}   
\makeindex

\begin{document}

  \def\apj{ApJ}
     \def\pasp{PASP}
     \def\pasj{PASJ}
     \def\araa{ARA\&A}
     \def\aap{A\&A}
     \def\apjl{ApJL}
     \def\apjs{ApJS}
     \def\mnras{MNRAS}
     \def\aj{AJ}
     \def\nat{Nature}
     \def\grl{GRL}
     \def\icarus{Icarus}
      \def\baas{BAAS}

\def\fref{\stackrel{\hspace{0.5ex}{\scriptscriptstyle\circ}}{f}
            \hspace{-0.8ex}}
\def\tauref{\tau^{\hspace{-0.9ex}^{\circ}}}
\def\nfref{n_{\rm f}^{\hspace{-0.9ex}^{\circ}}}
\def\nrref{n_{\rm r}^{\hspace{-0.9ex}^{\circ}}}
\def\nref{n^{\hspace{-0.9ex}^{\circ}}}
\def\vref{v^{\hspace{-0.9ex}^{\circ}}}
\def\alpharef{\alpha^{\hspace{-0.9ex}^{\circ}}}
\def\pref{p^{\hspace{-0.9ex}^{\circ}}}
\def\st{{\rm \hspace{0.1ex}\circ\hspace{-1.2ex}-}}
\def\dG{\Delta_{\rm f}^{\st}\hspace{-0.2ex}G}
\def\pst{p^{\st}}
\def\Vl{{V_{\ell}}}
\def\pabl#1#2{\frac{{\rm\partial} #1}{{\rm\partial} #2}}
\def\vgas{\vec{v}_{\rm gas}}
\def\vdreq{v^{\hspace{-0.8ex}^{\circ}}_{\rm dr}}
\def\xx{\vec{x}}
\def\ie{i.\,e.\ }
\def\eg{e.\,g.\ }
\def\DV{\Delta V_r}
\def\Vs{{V_{\rm s}}}
\def\Sr{S_{\!r}}
\def\vdrn{\vec{v}_{\rm dr}^{\hspace{-0.9ex}^{\circ}}}
\def\nr0{n_{r}^{\hspace{-0.9ex}^{\circ}}}
\def\jdiff{\vec{j}^{\rm diff}_i}
\def\er{\vec{e}_r}
\def\rhod{\rho_{\rm d}}
\def\nH{n_{\langle{\rm H}\rangle}}

\chapter[Cloud formation  in Exoplanetary Atmospheres]{Cloud formation  in Exoplanetary Atmospheres\label{ra_ch1}}

\author[Ch. Helling]{Ch. Helling}

\address{Space Research Institute, Austrian Academy of Sciences, Schmiedlstra{\ss}e 6, A-8042 Graz, Austria\\
Institute of Theoretical and Computational Physics, TU Graz, Petersgasse 16, Graz, Austria\\
Centre for Exoplanet Science,  University of St Andrews, St Andrews, KY16 9SS, UK}

\begin{abstract}
This invited review for young researchers presents key ideas on cloud formation as key part for  virtual laboratories for exoplanet atmospheres. The basic concepts are presented, followed by utilising a time-scale analysis to disentangle process hierarchies.  The kinetic approach to cloud formation modelling is described in some detail to allow the discussion of cloud structures as prerequisite for future extrasolar weather forecasts.
\end{abstract}


\body

\section{Introduction} 
The characterisation of exoplanets is a driving science goal for all future exoplanet space missions (PLATO, Ariel, LUVOIR/HabEx).  The essential task is to determine which explanet hosts an atmosphere. This is, in principle, the first step for future space travel. The detection of gaseous species, for example H$_2$O \cite{2020MNRAS.497.5155W,2020AJ....160..280C,2021A&A...646A.168C}, AlO \cite{2020A&A...639A...3C}, CrH \cite{2021A&A...646A..17B}, and also ions like Ca$^+$ and Fe$^+$ \cite{2019A&A...628A...9C,2020A&A...640C...6C}, point to the presence of an atmosphere. Specific molecules can be used to trace specific atmospheric processes,  for example is H$_3^+$ an Auroral tracer for Jupiter, and hence, a fingerprint for the presence of a planetary magnetic field\cite{2018NatAs...2..773S,2020RvMP...92c5003M}.  The possible detection of HCN \cite{2017pre8.conf..345H} in particular on the nightside of an exoplanet \cite{2021MNRAS.502.6201B} may point to the occurrence of lightning in the planet's atmosphere. Lightning, that has a multitude of observable signatures \cite{2016MNRAS.461.3927H} is not only a sign for the presence of an atmosphere, but also for the presence of clouds. The presence of clouds in exoplanet atmosphere has so far been inferred indirectly as the 'dark matter' that is needed to reproduce the less-then-expected molecular absorption signatures; often by prescribing cloud properties as part of retrieval approach \cite{2018AJ....155...29W,2020MNRAS.497.4183B,2021ApJ...913..114W}.  Importantly, the presence of clouds offers another proof for the presence of an atmosphere on an extrasolar planet. 

Clouds  hamper efforts to characterise exoplanet atmospheres by remote sensing  from Earth as clouds  block the view into the deeper atmosphere. Young rocky planets  may offer additional potential for confusion if their  atmosphere exhibit mineral clouds which may be mistaken as a rocky/sandy surface. For example, it may not be immediately clear if the albedo is the result of a mineral cloud layer or a rocky surface since the reflectance of sand is practically wavelength independent at  $\lambda > 0.6\mu$m \cite{2016IJAsB..15...45B}.

Atmospheric clouds actively influence their environment by radiative heating and cooling, by depleting and/or enriching the local gas phase, and by locking up elements and potentially transporting elements though the atmosphere. Therefore, in order to determine the presence and abundance of key molecules, atoms or ions  as well as the overall element abundances that allow links to planet formation and evolution, cloud formation must be understood in enough detail. Cloud particles or {\it aerosols} may further enable the formation of biomolecules under the influence of charges \cite{2014IJAsB..13..165S} or  could support the metabolism of aerial  life forms through chemolithoautotrophy. The concept of extraterrestrial aerial biospheres as attracted renewed attention with respect to the Venusian cloud layers\cite{2021Univ....7..172S}. The formation of clouds is key to all these important aspects such that we strive to understand how clouds form in atmosphere of a  wide chemical composition in the first place \cite{2019AREPS..47..583H}.   Lastly, cloud formation occurs within a thermodynamic setting which is modelled at different degrees of complexity within the exoplanet community\cite{2021JGRE..12606655G,2021MNRAS.502.2198T}.


\begin{figure} 
\centerline{\includegraphics[width=10cm,angle=-90]{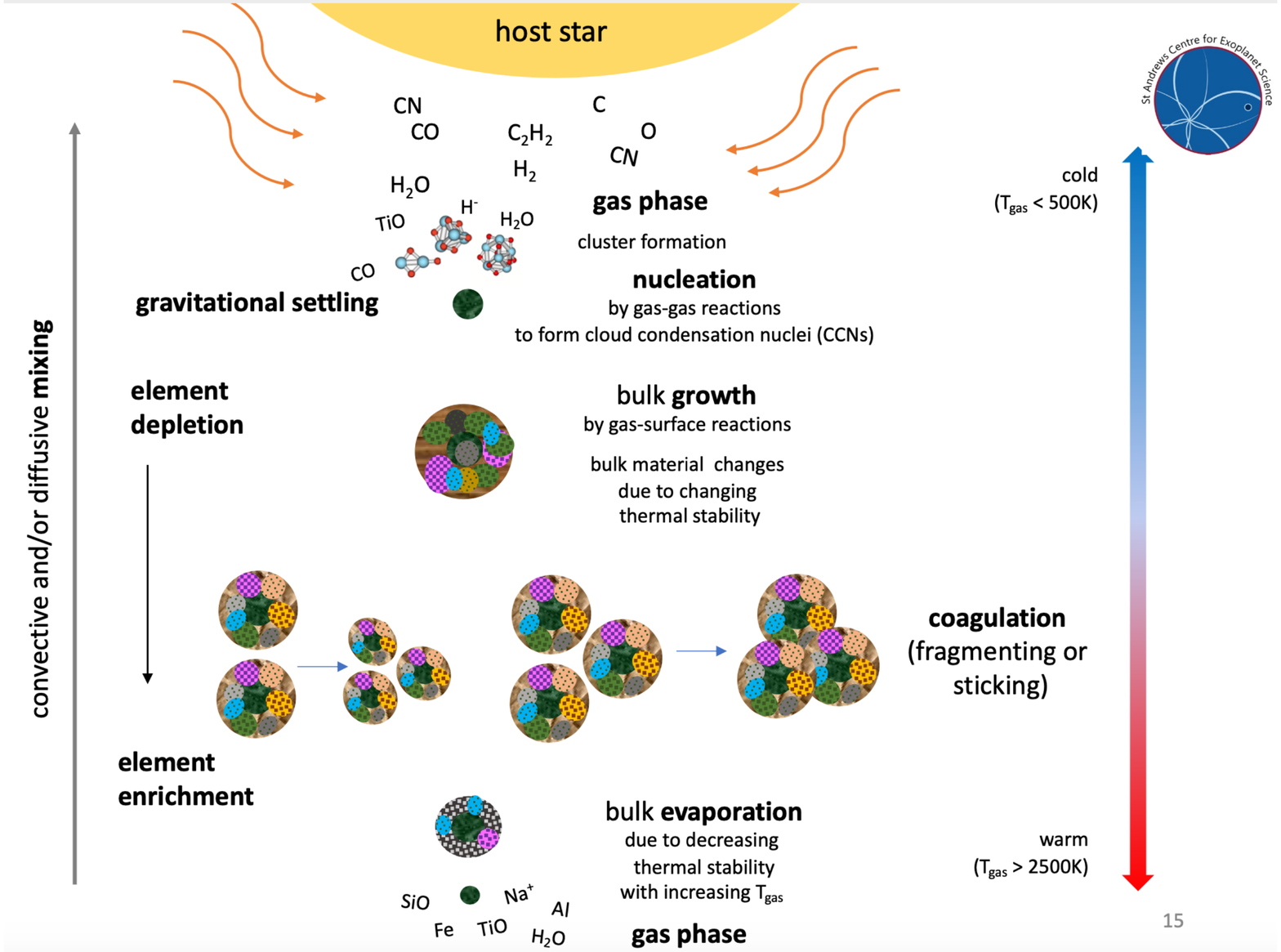}} \caption{Cloud formation processes:  First, condensation seeds  form through gas-gas collisions either directly in the gas phase or triggered by photo-chemistry. Secondly, these condensation seeds grow through gas-surface reaction which grow the bulk of the cloud particle mass. These cloud particles gravitatiallly settle  into the deeper, denser and also warmer atmospheric layers. They may coagulate and form larger particles or they may shatter through particle-particle  collisions. At higher temperature, the cloud particles evaporate. Convective or diffusive mixing may cause element replenishment. } \label{fig:cloudfromsketch} \end{figure}

\section{A basic concept for cloud formation}\label{s:ccf}

Cloud formation on Earth most commonly starts with the condensation of gaseous water (H$_2$O[g]) that has just reached supersaturation, i.e. the supersaturation ratio $S = p(T_{\rm gas})_{\rm H2O}/p(T_{\rm gas})_{\rm vap, H2O} = 1.001$ ($p(T_{\rm gas})_{\rm H2O} $ -- local partial pressure of H$_2$O[g], $p(T_{\rm gas})_{\rm vap, H2O}$ -- local saturation vapour pressure of H$_2$O[l])) on a number of pre-existing cloud condensation nuclei (CCNs). The thermal stability of the condensate (liquid water H$_2$O[l])  is the driving principle and applies not only to the condensed phase (liquid or solid) but also to the molecular clusters that will eventually form the CCN. The governing processes of cloud formation are visualised in Fig.~\ref{fig:cloudfromsketch} and start with formation of CCNs ({\bf nucleation}) in the most general case.  CCN are the result of the phase transition form the gas to the solid phase and provide the first surface that forms from the gas phase. These CCN enable the {\bf bulk growth} of already thermally stable materials since a surface reaction requires less energy than  the gas-gas interactions that lead to the formation of a CCN. Once these particles (or droplets or aerosols) have formed they fall through the atmosphere, i.e. they gravitationally settle.  The {\bf gravitational settling} is determined by the force equilibrium between gravitational and frictional force. The particles will experience different frictional force regimes (turbulent, laminar, viscous, free molecular flow) depending on their size, their density and the ambient gas density.  The fall through the atmosphere, in fact, any motion of the cloud particles within the atmosphere, causes the cloud particle to encounter changing thermodynamic conditions.  This can have two different effects:  Either additional material become thermally stable and condenses onto the cloud particle  such that the cloud particle's composition increases its mixture of materials, or, the temperatures are too high and material {\bf evaporate} such that the cloud particle's composition decreases its mixture of materials. It is important at this point to understand that the condensation and the evaporation of materials is linked to the gas-phase abundances in that condensation consumes gas constituents and evaporation provides gas constituent to the atmospheric environments. This simple fact is termed {\bf element conservation}, and it can be concluded that cloud particle moving through the planetary atmospheres transport elements around, i.e. provide a mechanism of element mixing in exoplanet atmospheres.

Once cloud particle are present, they can undergo alterations through changing thermodynamic conditions of their environment and by collision processes between existing particle. Collisional process may lead to charging cloud particle (called triboelectric charging), and also to the formation of larger cloud particle through sticking ({\bf coagulation}) or to cloud particle destruction through shattering ({\bf fragmentation}).  The outcome of the collisional process (coagulation vs fragmentation) is predominantly determined by the relative velocity between the cloud particles. Within these two groups, the number of  processes that may occur is large since the collisional partner may have different masses \cite{2010A&A...513A..56G}. A large body of work  has been conducted in the planet-forming disk community on this topic.

The more exoplanetary environments we discover, the further processes may need to be considered to study cloud formation and its feedback on exoplanet atmospheres.  Charge processes are such a class of processes that attract an increasing amount of attention since they are triggered by the host star's radiation field in the observable part of the atmosphere and they may trigger processes than enable to formation of pre-biotic molecules either on cloud particles are through lightning. Additionally, ion-ion reaction have shown to play an essential role for the formation of CCNs on Earth in the higher atmospheric regions\cite{2016JGRA..121.8152S}. A basic understanding of cloud formation will help to address some of these challenges\cite{2019JPhCS1322a2028H}, but not all.

\section{A time-scale analysis}\label{s:timescl}

It will be helpful to run though a time-scale analysis of the basic cloud formation  processes. Figure~\ref{fig:timescales} compares how long which processes may take inside the atmospheric environment of a warm gas giant of a global temperature of T$_{\rm eff}=1400$K and a surface gravity of $\log$g = 3.0 (cm\,s$^{-1}$). The same set of global parameters may also describe a young brown dwarfs.

\begin{figure} 
\label{fig:timescales}
\centerline{\includegraphics[width=12cm]{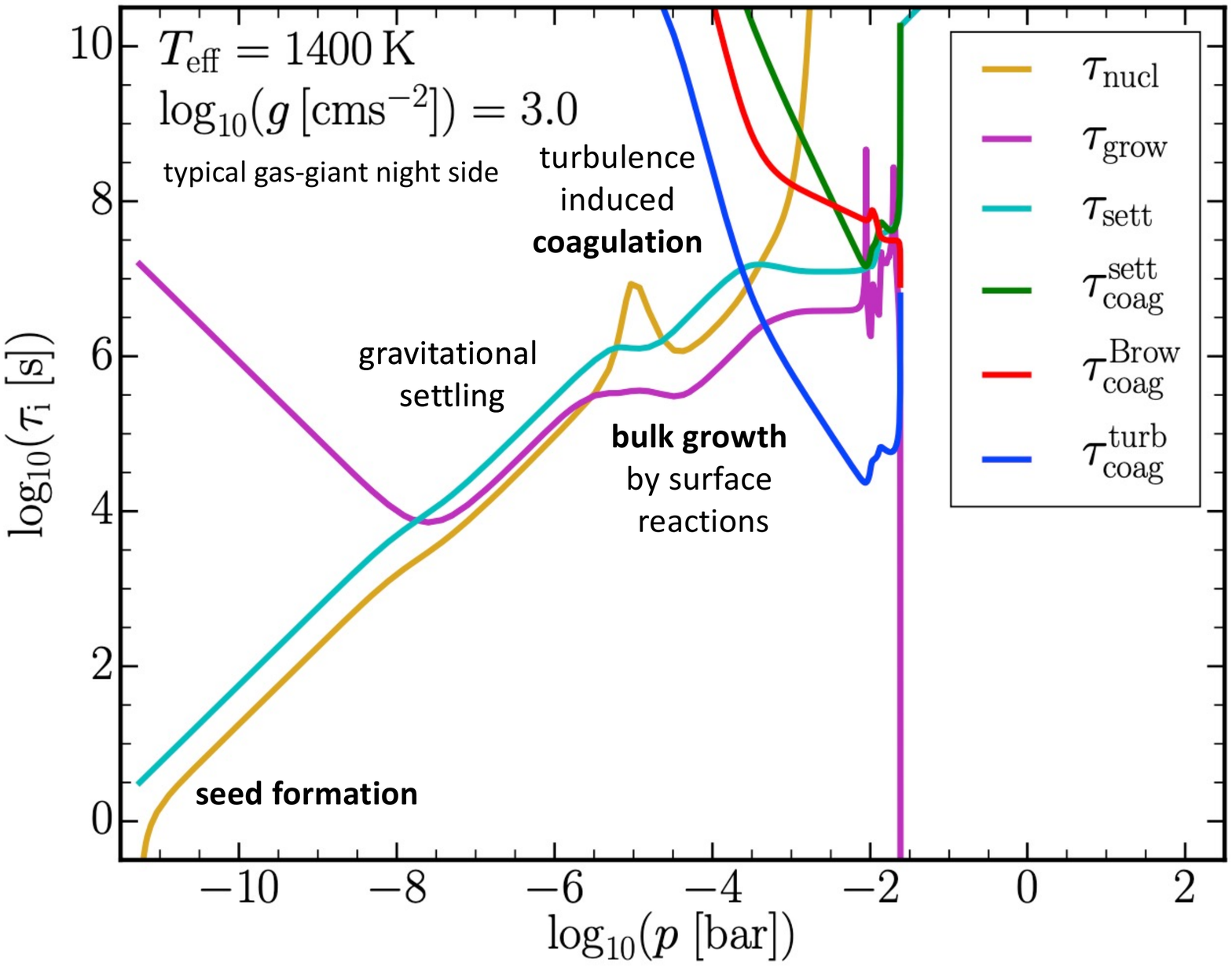}}
\vspace*{-0.5cm}
 \caption{Time scale comparison of basic cloud formation processes for a 1D {\sc Drift-Phoenix}\cite{2009A&A...506.1367W} atmosphere for a cloud-forming gas giant  [courtesy to D. Samra].} \label{fig:timescales} \end{figure}
 
 \noindent
 The individual time-scales are defined as follows:\\
 -- The nucleation time scale, $\tau_{\rm nuc}=(\rho L_0)/J_*$, is the time that it takes to form a number of CCNs ($\rho L_0$) through a certain nucleation rate ($J_*$). Here, the formation of TiO$_2$ CCNs is considered.\\
 -- The growth time scale, $\tau_{\rm grow}= \langle a\rangle / \chi^{\rm net}$, is the time that it takes to grow a cloud particle of a mean particle size ($ \langle a\rangle$) with a certain net surface growth velocity ($\chi^{\rm net}$). \\
-- The settling time scale, $\tau_{\rm set}= H_{\rm p}(T_{\rm gas})/v_{\rm dr}( \langle a\rangle)$,  is the time that it takes a particle of a certain mean size, $ \langle a\rangle$, to fall with a certain drift velocity ($v_{\rm dr}( \langle a\rangle)$) \cite{2003A&A...399..297W} over a length of the pressure scale height ($H_{\rm p}(T_{\rm gas}$). \\
-- The time scales that characterise coagulation are $\tau^{\rm sett}_{\rm coag}$,  $\tau^{\rm Brow}_{\rm coag}$, and  $\tau^{\rm turb}_{\rm coag}$, and describe the time it takes for two particle to collide if they move with their settling velocity ($\tau^{\rm sett}_{\rm coag}$), their Brownian motion ($\tau^{\rm Brow}_{\rm coag}(T_{\rm gas})$) or with some turbulent velocity ($\tau^{\rm turb}_{\rm coag}$).

Figure~\ref{fig:timescales}  shows in which atmospheric regions cloud formation (nucleation, growth) or processing (coagulation) may occur. Nucleation will dominate  over a large portion of the atmospheric pressure range for $p_{\rm gas} = 10^{-11}\,\ldots\,10^{-5.5}$bar. The surface growth kicks in at  $p_{\rm gas} \approx  10^{-8}$bar efficiently, but dominates over nucleation only in deeper atmospheric layers $p_{\rm gas} = 10^{-5.5}\,\ldots\,10^{-3.5}$bar. Nowhere in this particular atmosphere do the cloud particles fall faster than they grow, hence $\tau_{\rm set} > \tau_{\rm grow}$ everywhere. Collisional interaction between existing cloud particles become of interest where $p_{\rm gas} >  10^{-3}$bar where $\tau^{\rm turb}_{\rm coag} \ll ( \tau_{\rm grow}\, \mbox{and} \,\tau_{\rm nuc})$. It is interesting to note that collisional particle-particle processes due to Brownian motion and gravitational settling are negligible compared to turbulence driven collisional interactions for atmospheres where the turbulence cascade is driven by convective instabilities\cite{2004A&A...423..657H}. Hence, turbulence driven particle-particle collision may lead to coagulation or to fragmentation of the respective cloud particles. Further insight requires to model the processes in detail.

\section{Cloud formation modelling}  

An extensive description of cloud formation modelling in exoplanet and brown dwarf environments was provided by Helling \& Fomins\cite{2013RSPTA.37110581H}.  Gail \& Sedlmayr\cite{2013pccd.book.....G} (Chapters 6, 9 to 14) present a text book with detailed explanations of basic  principle, hence applicable to the wide range of exoplanets. We refrain here from referring to more recent papers that present the same material in their own words.

\begin{figure} 
\centerline{\includegraphics[width=12cm]{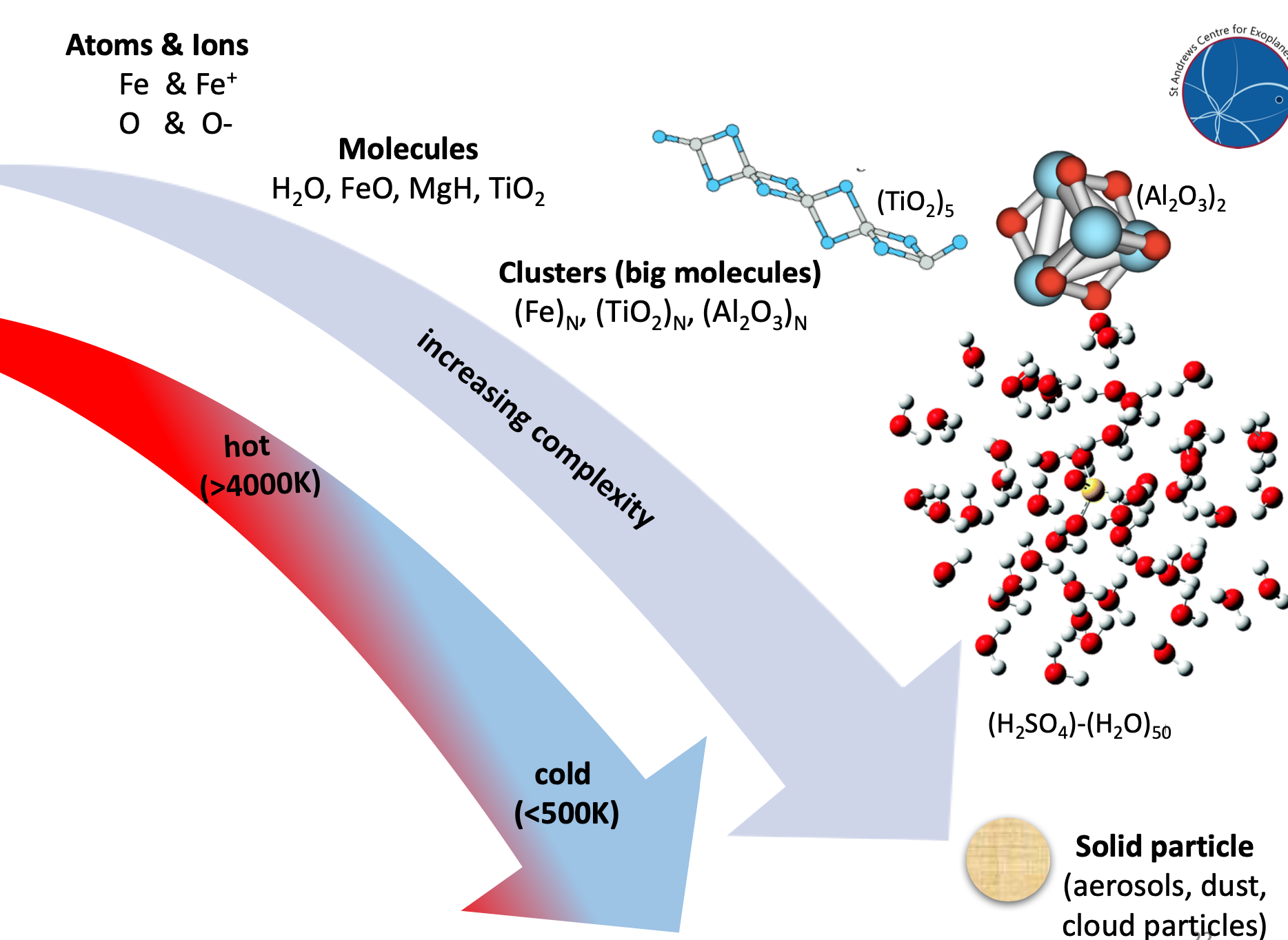}}
 \caption{The formation of condensation seeds leads to an increasing structural complexity, from the state of an unordered gas to a particle of a highly ordered structure.} \label{fig:incrcompl} \end{figure}

\subsection{Modelling the formation of condensation seeds}

To  model  the formation of condensation seeds means to  describe the formation of subsequently larger molecules the biggest of which are called clusters (see Fig.~\ref{fig:incrcompl}). Ideally, the complete chemical path from the gas phase to a solid particle could be modelled. This, however, is challenging since not only one path may exists and material data for all possible cluster sizes as well as their isomers will be required. If it is possible to identify one such chemical path that dominates the phase transition, then one can identify characteristic properties like a critical cluster and a bottle neck reaction.

Let us aim to incorporate as much microphysical knowledge as we possibly can into our model for condensation seed formation.
The kinetic, steady-state approach describes the temporal evolution of the cluster size distribution function, $f(N, t)$,  as \citep{2013RSPTA.37110581H},
\begin{equation}
\label{eq:master}
\frac{df(N,t)}{dt} = \sum_{i=1}^I J^c_i(N,t) -  \sum_{i=1}^I J^c_i(N+i,t).
\end{equation}
where $f(N)$ is the number density of a molecular cluster containing $N$ $i$-mers  that we are interested in.
$J^c_i(N,t)$ is the effective flux (or transition rate) for
the growth of the cluster of size $N-i$ to size $N$. This 
flux through cluster space is 
\begin{equation}
\label{eq:clflux}
J^c_i(N,t) = \sum^{\rm R_i}_{\rm r_i -1} 
            \left(\frac{f(N-i, t)}{\tau_{\rm gr}(r_i, N-i, t)} - \frac{f(N, t)}{\tau_{\rm ev}(r_i, N, t)}\right), 
\end{equation}
summing over all chemical reactions $r_i$ in
which an $i$-mer is involved.  $\tau_{\rm gr}(r_i,
N-i, t)$ is the growth time by reaction $r_i$ leading
from cluster for size $N-i$ to cluster size $N$. $\tau_{\rm ev}(r_i, N, t)$ is the evaporation time leading from size $N$ to size $N-i$,
\begin{eqnarray}
\label{eq:taugr}
\frac{1}{\tau_{\rm gr}(r_i, N-i, t)} &=& A(N-i)\, \alpha(r_i, N-i) v_{\rm rel}(n_{\rm f}(r_i), N-i)\, n_{\rm f}(r_i)\\
\label{eq:tauev}
\frac{1}{\tau_{\rm ev}(r_i, N, t)}&=&A(N)\, \beta(r_i, N)\, v_{\rm rel}(n_{\rm r}(r_i), N)\, n_{\rm r}(r_i).
\end{eqnarray}
$n_{\rm f}(r_i)$ and $n_{\rm r}(r_i)$ are the number density of the molecule of the growth (forward) process and of the evaporation (reverse) process for reaction $r_{\rm i}$\citep{1998A&A...337..847P}. $A(N)$ is the surface of the cluster of size $N$. 
 $v_{\rm rel}$ is the average relative velocity between the
growing/evaporating TiO$_2$ molecule and the (TiO$_2$)$_{\rm N}$ cluster. It is defined as the thermal velocity 
\begin{equation}
v_{\rm rel} = \sqrt{ \frac{k_B T}{2\pi} \left(\frac{1}{m_{\rm N}} + \frac{1}{m_{\rm TiO_2}}\right)}.
\end{equation}
 $\alpha(r_i, N-i)$ and $\beta(r_i, N)$ are the reaction efficiency for growth and evaporation via reaction $r_{\rm
  i}$. Note that $\alpha$ is also called the sticking probability.   
The growth and evaporation efficiency
coefficients, $\alpha(r_i, N)$ and $\beta(r_i, N)$, are often unknown for the different cluster sizes $N$. 3D Monte-Carlo molecular-dynamic-like  simulations can help to study the effect of uncorrelated reaction efficiencies\cite{2021arXiv210804701K}.

Following Patzer et al. (1998), a {\it reference  equilibrium state}  is introduced to be able to solve Eq.~\ref{eq:master} with Eqs.~\ref{eq:taugr} and~\ref{eq:tauev}. Patzer et al. (1998) show in their Appendix A that if the temperatures of all components are equal, the supersaturation ratio of a cluster of size $N$ with respect to the bulk is $S_{\rm N}= (S_1)^{\rm N}$. The reference equilibrium state is therefor characterised by  phase equilibrium between monomers and between the clusters and the bulk solid, plus simultaneous chemical equilibrium in the gas phase, plus thermal equilibrium (i.e. all components have the same temperature). In such local thermodynamic equilibrium (LTE) between the gas phase and the clusters the principle of detailed balance holds for a single microscopic growth process and its respective reverse, i.e. the evaporation process. This implies that under the condition of detailed balance, $\fref(N-1)/\tau_{\rm gr}(N-1)=\fref(N)/\tau_{\rm ev}(N)$, which allows to express the evaporation rate by the growth rate. 
$\fref(N-i)$, $\fref(N)$, $\nfref(r_i)$, and $\nrref(r_i)$ are the
equilibrium number densities for the clusters 
and the monomers.
The law of mass action described  the link between the 
equilibrium particle densities and Gibbs free energies,  
\begin{equation}
\label{eq:lma}
\left( \frac{\fref(N-i)\, \nfref(r_i)}{\fref(N)\,\nrref(r_i)}\right) = 
 \exp \left(
      \frac{\dG(r_i, N, T_{\rm d}(N))}
           {R\,T_d(N)}
      \right).
\end{equation}
$\dG(r_i, N, T_{\rm d}(N))$ [kJ mol$^{-1}$] is the Gibbs free energy
of formation. It is calculated from the standard molar Gibbs free
energy of formation of all reaction participants at the temperature
$T_{\rm d}(N)=T_{\rm gas}$ as
\begin{eqnarray}
\label{eq:dG1}
\nonumber
\dG(r_i, N, T_{\rm d}(N)) &=& \dG(N,  T_{\rm d}(N)) - \dG(N-i,  T_{\rm d}(N))\\
                          &+& \dG(n_{\rm r}(r_i), T_{\rm d}(N))\\
                          \nonumber
                          &-&  \dG(n_{\rm f}(r_i), T_{\rm d}(N)).
\end{eqnarray}
In LTE, the equilibrium cluster size distribution is therefore expressed by a Bolzmann
distribution,
\begin{equation} 
\label{eq:fref}
\fref(N) = \fref(1)\,\exp\left(-\frac{\Delta G(N)}{RT}\right),
\end{equation} 
with $\fref(1)$ the equilibrium density of the monomer (for example TiO$_2$, SiO, NaCl). 
$\Delta G(N)$ is the free energy change due to the formation of a
cluster of size N from the saturated vapour. It is related to the
standard molar Gibbs free energy of formation of the N-cluster
$\dG(N)$ by
\begin{equation}
\label{eq:dG}
\Delta G(N) = \dG(N, T) + RT \ln\left(\frac{p_{\rm sat}(T)}{\pst}\right) - N\,\dG_1(s, T).
\end{equation}
$\dG_1(s),T$ is the standard molar Gibbs free energy of the formation of the
solid phase, and $\pst$ is the pressure of the standard state. Most
often, $\pst$ is the atmospheric pressure on Earth at which
$p_{\rm sat}$ and $\dG(N, T)$ were measured. {\it The right hand side of
Eq.~\ref{eq:dG} contains now quantities which can be determined from
lab experiments or quantum-chemical calculations.}

Note that the saturation vapour pressure is defined for an infinitely extended, planar surface but clusters do have a curvature.  This surface curvature does affect the thermal stability of the cluster which was well recognised as a challenge within the astronomical community studying dust formation in late-type stars\cite{1996ASPC...96...69G}. In order for clusters to form,  the gas needs to be considerably supersaturated (or super-cooled) in order to enter the regime of thermal stability for clusters. Furthermore, no assumptions were necessary with respect to size-independent surface tensions. Only a classical nucleation approach (droplet model) requires the consideration of the so-called Kelvin effect. The Kelvin effect is therefore used to include some representation of the curvature of the clusters/particles for the evaporation process. The use of thermodynamic cluster data in combination with  the assumption of detailed balance conveniently side-steps calculating evaporation such that the concept of  surface tension is not required to describe the effect of surface curvature on thermal stability.

The final step is  to derive an expression that enables to calculate the rate at which cloud condensation nuclei form. For this, we assume a stationary flow through the cluster space to occur. The slowest reaction will determine the flux through the cluster space during  a chain of reaction. This bottle-neck reaction will lead to the formation of the critical cluster, $N_*$ (Eq.~4.15 in Helling \& Fomins 2013) , which is the least stable cluster, and once it has formed, constructive cluster growth processes will dominate. This means that the nucleation rate, $J_*$, is determined by the quantities of the critical cluster.  For a homogeneous, homomolecular process in thermal equilibrium one finds\cite{2013RSPTA.37110581H}
\begin{equation}
J_* = \frac{\fref(1)}{\tau_{\rm gr}(1, N_*)}\,Z(N_*) \cdot \exp\Big((N_*-1) \ln S(T) - \frac{\Delta G(N_*)}{RT}\Big),
\end{equation}
with $Z(N_*)$ the Zeldovich factor (Eq.~4.16 in Helling \& Fomins 2013).

\begin{figure} 
\centerline{\includegraphics[width=\textwidth]{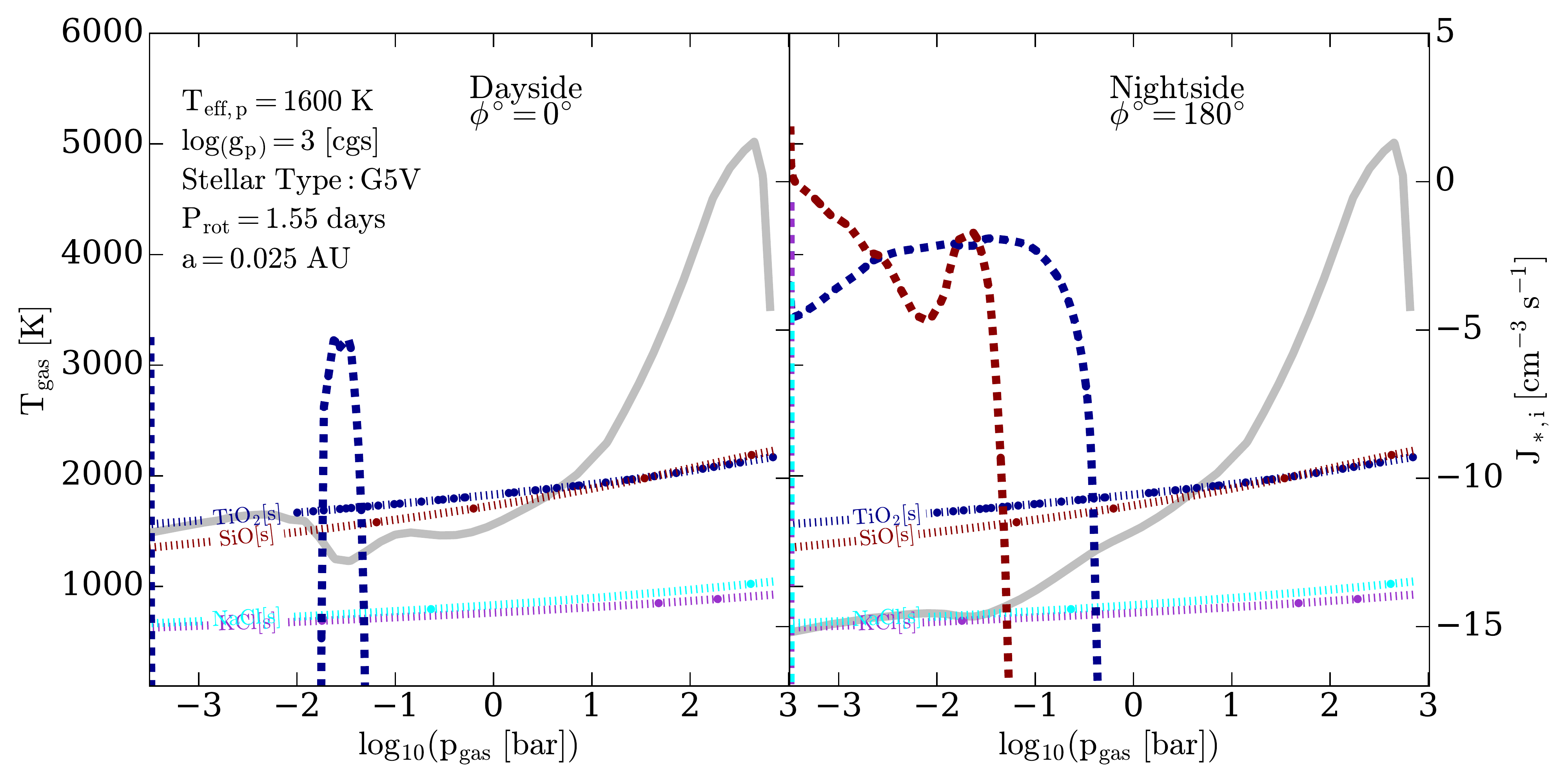}}
\vspace*{-0.5cm}
 \caption{Comparing the location of  the homogeneous, homomolecular nucleation rate  for TiO$_2$ (dashed dark blue line), SiO (dashed brown line), NaCl (not forming here), and KCl (not forming here) and  the supersaturation ratio $S_{\rm TiO2[s], SiO[s], NaCl[s], KCl[s] }=1$ in the $(T_{\rm gas}, p_{\rm gas})$-plane. The nucleation rates are calculated for  $(T_{\rm gas}, p_{\rm gas})$-profiles (grey solid lines)  of the dayside (left) and the nightside (right) for a tidally locked giant gas planet. The formation of CCNs is only possible if the gas is considerably colder than where thermal equilibrium (S=1) occurs.  [courtesy to D. Lewis ]} 
 \label{fig:Teffpl1600KSsat}
 \end{figure}

Figure~\ref{fig:Teffpl1600KSsat} demonstrates how the location in the $(T_{\rm gas}, p_{\rm gas})$ plane  compares for  results for the  homogeneous, homomolecular nucleation rates for TiO$_2$ and SiO (dashed lines, material data similar to Lee et al. 2018 (\cite{2018A&A...614A.126L})) and the location of thermal stability where the supersaturation $S=1$ for each of the solids (dotted lines). The modified nucleation theory is used to derive $N_*$ and  cluster data to derive the required surface tension for TiO$_2$ clusters is applied. Recall that no phase transition is possible in thermal equilibrium which is represented by $S=1$ here. Hence,  material will evaporate of $T>T(S=1)$ and it will be thermally stable for $T\leq T(S=1)$. Therefore, CCN formation can only occur if  $T_{\rm gas}\ll T(S=1)$.

The nucleation rates are calculated for pre-scribed 1D atmosphere profiles  (solid grey line) for the dayside (left) and the nightside (right) of a tidally locked gas giant planet (for details see Sect.~\ref{s: results}). Figure~\ref{fig:Teffpl1600KSsat}  demonstrates that nucleation only occurs if the local gas temperature is sufficiently lower than $T_{\rm gas}(S=1)$. The nucleation rate is lower on the dayside (left) where the local gas temperature (solid grey curve) decreases just below the $T_{\rm gas}(S=1)$ compared to the nightside where a substantial supercooling below $T_{\rm gas}(S=1)$ occurs. 
The dayside features $T_{\rm gas}\approx T_{\rm gas}(S_{\rm TiO2}=1)$ at $p_{\rm gas}\approx 10^{-2.5}$bar where consequently no TiO$_2$-nucleation occurs. The gas temperatures are too high for NaCl and KCl to nucleate. 

\subsection{Modelling the bulk growth of cloud particles}

Section~\ref{s:timescl} has shown that seed formation is the fastest process until the bulk growth supersedes it in higher density and warmer atmospheric regions. This section summarises a method that allows to efficiently simulate the bulk growth of cloud condensation nuclei through surface reaction. The supersaturation ratio distinguishes now the constructive material growth ($S>1$) from the destructive material evaporation ($S<1$).

Firstly, a conservation equation (the master equation) for the cloud particle size distribution, 
$f(V)$ [${\rm cm^{-6}}$],  in  volume space is formulates as
\begin{equation}
\label{eq:master}
    \frac{\partial}{\partial t}\left(f(V) {\rm d}V \right) + \nabla \cdot \left(v_{\rm d} (V)  f(V){\rm d}V \right) = \sum_{\rm k} R_{\rm k} {\rm d}V.
\end{equation}
Equation~\ref{eq:master} can be solved directly by spectral binning methods\cite{2018ApJ...860...18P,2019ApJ...876L...5K} or indirectly by defining moment of this equation.
In order to derive a comprehensive set of equations that allow to derive cloud properties like particle sizes and material compositions,  moments, $L_{\rm j}$, are defined  as
\begin{equation}
	\rho L_{j}=\int_{V_{\rm l}}^{\infty}f(V)V^{j/3}{\rm d}V,
	\label{equ:Moment_definition}
\end{equation}
which converts Eq.~\ref{eq:master} into a set of moment equations for $j=0, 1, 2, \ldots$

\begin{equation}
	\frac{\partial}{\partial t}(\rho L_{j})\ +\ \nabla \cdot ({v}_{\rm d}\rho L_{j}) = \int_{V_{\rm l}}^{\infty} \sum_{k} R_{k}V^{j/3} {\rm d}V.	\label{equ:mommaster}
\end{equation}
The cloud particle velocity $v_{\rm d}(V)$ can be expressed in terms of the local gas velocity and the relative velocity between a cloud particle of volume $V$ and the surrounding gas phase of density $\rho$ as $v_{\rm d} (V) ={v}_{\rm gas} + {v}_{\rm dr}(V) $. The frictional interaction of a cloud particle with the surrounding atmosphere does depends on its mass, size and the atmosphere density and different flow regimes may results \cite{2003A&A...399..297W}. In a subsonic free molecular flow (large Knudsen numbers ${\rm lKn}$), and 1D plane parallel geometry ($z$ direction only), Eq.~\ref{equ:mommaster} becomes
\begin{multline}
	\frac{\partial}{\partial t}(\rho L_{j})\ +\ \nabla \cdot (v_{\rm gas}\rho L_{j}) = \\ V_{\rm l}^{j/3} J_*\ +\ \frac{j}{3} \chi^{\rm net}_{\rm lKn}\rho L_{j-1}\ +\xi_{\rm lKn}\rho_{\rm d} \frac{\partial}{\partial z}\left (\frac{L_{j+1}}{c_{T}}\right).
	\label{eq:mommaster_largeKn_planeparralel}
\end{multline}
The first term on the r.h.s describes the nucleation process, the second the surface growth and the third the effect of gravitational settling which transports the cloud particles with their relative velocities (raining particles). The lower integration boundary $V_{\rm l}$ represents the size of the CCNs and differs for the different nucleation species (e.g. TiO$_2$, SiO, NaCl, KCl). $\chi^{\rm net}_{\rm lKn}\sim n_{\rm r}^{key} (1- 1/S_r)$ is the total growth velocities for all materials $s$ and defined in Eq.~5.43 in Helling \& Fomins 2013 (\cite{2013RSPTA.37110581H}), and $\xi_{\rm lKn}= (3/4\pi)^{1/3} (\sqrt{\pi}/2)g$. $r$ annotates the chemical surface reactions. Various reactions $r$ will contribute to the growth of a specific material $s$. The supersaturation is here defined with respect to a certain reaction $r$. For more in-depth details, please refer to our original papers as cited.  Once the moment equations (Eqs.~\ref{eq:mommaster_largeKn_planeparralel}) are solved, the moments $\rho L_{\rm j}$ (j=0, 1, 2, 3, ...) can be used to derive cloud properties:\\
-- cloud particle number density: $n_{\rm d} = \rho L_0$\\
-- mean particle size $\langle a\rangle =  \sqrt[3]{3/4\pi}\cdot (L_1/L_0)$ (Appendix A in  (\cite{2020A&A...641A.178H}))\\
-- mean particle surface  $\langle A\rangle  =  \sqrt[3]{36\pi} L_2/L_0 = A_{\rm cloud}^{\rm tot}/n_{\rm d}$\\
-- mean particle volume  $\langle V\rangle  =  L_3/L_0 = \sum L_3^{\rm s}/L_0$ with $V_{\rm s}=L_3^{\rm s}/L_0$.

Average quantities represent an ensemble rather than individual particles. What may sound like a shortcoming is, in fact, a strength. By utilising higher order moments ($L_3$ and $L_2$) it is possible to derive an ensemble average more suitable, for example, for opacity calculation for which the surface area is more suited (opacity calculation)\cite{2020A&A...641A.178H}.

\begin{figure} 
\centerline{\includegraphics[width=\textwidth]{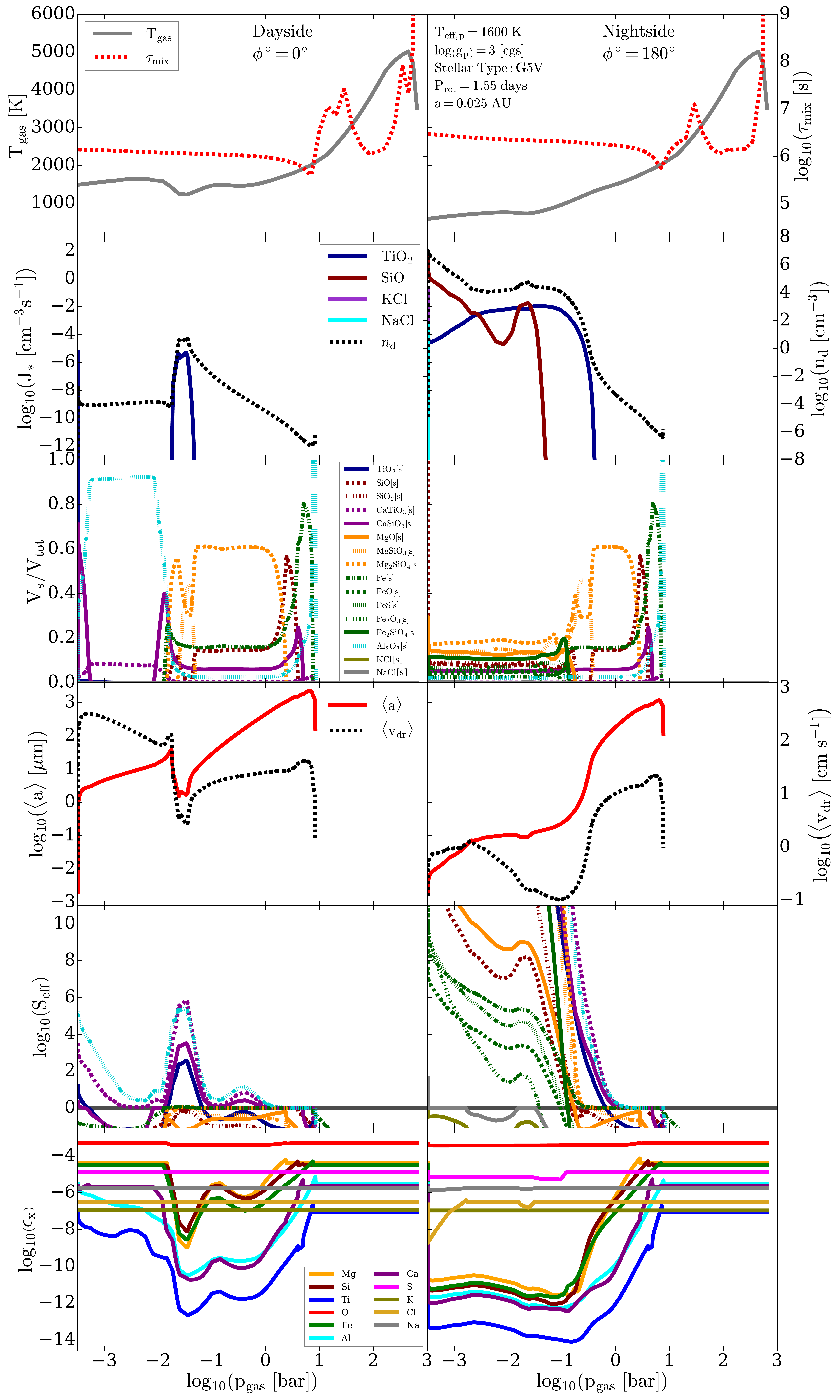}}
\vspace*{-0.5cm}
 \caption{The cloud profiles for the substellar ($\phi=0^o$, dayside) and the antistellar ($\phi=180^o$, nightside)  point of a tidally locked gas-giant orbiting a G-type host star. The gas-giant would have an effective temperature of T$_{\rm eff}$=1600K and a log(g)=3.0 for a set solar element abundances. The 1D profiles are extracted form a 3D GCM simulation by  Baeyens et al. (2021)\cite{2021MNRAS.tmp.1277B}. [courtesy to D. Lewis ].} 
 \label{fig:Teffpl1600K}
 \end{figure}

\section{The resulting cloud structure,\\ the prerequistis for extrasolar weather forecast}\label{s: results}

The numerical solution of the cloud formation model described in the previous sections in 1D and for a stationary cloud in an atmosphere that is well-mixed by convection allows to discuss the principle vertical cloud structures. In addition to the cloud model described, the element conservation has to be solved for each element involved and the chemical composition of the gas phase calculated. The cloud model in combination with the gas-phase chemistry calculation can be used as virtual laboratory in order to explore cloud details like the number of cloud particles forming, their material composition and the  expansion of the cloud for a set of  local gas temperatures, gas pressurse and the vertical mixing efficiencies ($\tau_{\rm mix}$). Here we compare the 1D vertical cloud structure at the dayside (substellar point) and the nightside (antistellar point) in Fig.~\ref{fig:Teffpl1600K} for a tidally locked, moderately warm giant gas planet that orbits a G-type host star at an orbital distance of 0.025 AU with a rotational period of 1.55 days. The same approach can be utilised to study many 1D profiles spanning the full planetary globe by extracting them from 3D GCMs in order to  provide a first-order exploration of cloud properties and their distributions for different extrasolar planet atmospheres structures
\cite{2021A&A...649A..44H}, hence, their local weather.

The local gas temperature determines the chemical composition of the gas phase and hence, the inset of the seed formation process. We note that the dayside of our model exoplanet in Fig.~\ref{fig:Teffpl1600K} has a gas temperature well above 1300K in the uppermost atmosphere where $p_{\rm gas}<10^{-2}$bar. Consequently, only the nightside allows for an efficient formation of seed particles of TiO$_2$ and SiO over 3 orders of pressure magnitude (2nd row, right, solid lines, $J_*$), whereas the region of efficient seed formation remains confined to a very narrow temperature interval on the substellar point  (2nd row, left, solid lines). Only TiO$_2$ (blue) contributes here to form cloud condensation nuclei. The nightside is cool enough that SiO  (brown) contributes considerably to the formation of cloud condensation nuclei in the uppermost, cooles layers of our  computational domain. The number density of cloud particle, $n_{\rm d}$,  is a direct consequence of the seed formation process. Figur~\ref{fig:Teffpl1600K} (2nd row,  dotted lines) furthermore demonstrates that $n_{\rm d}$ expands into  higher pressure regions compared  to the nucleation rate. That shows that cloud particles have fallen (gravitational settling) into the deeper atmosphere where seed formation becomes thermally impossible. As particles fall through the atmosphere, they encounter different chemical gas compositions as result of the changing  thermodynamic conditions (top panels, solid black lines). Consequently, the material composition of the cloud particles changes throughout the atmosphere (3rd row, material volume fractions $V_{\rm s}/V_{\rm tot}$). The very top layers are made of the cloud nucleation species TiO$_2$ and SiO which lose in importance where other, more complex materials condense and compete for the elements that they are made of. The middle portion of the cloud layer is made of a wide mix of Mg/Fe/Si/O materials peppered with materials that contain elements of lesser (solar) abundances like Ca or Al. The warmest parts of the clouds are made of high-temperature condensates like Fe[s], Al$_2$O$_3$[s], CaTiO$_3$[s]. 

The detailed material composition in the optically thin part of the cloud ($p_{\rm gas}<10^{-3}$bar) differs considerably between the day- ($\phi=0^o$) and the nighside ($\phi=180^o$) reflecting the very different local temperatures. A very low nucleation rate at $p_{\rm gas}<10^{-3}$bar causes a small number of cloud particles made of TiO$_2$[s], CaSiO$_2$[s], Al$_2$O$_3$[s] and CaTiO$_3$[s] to grow to considerably sizes  of 10 $\mu$m from these rarified gases. A considerably larger number of cloud particles are able to form on the nightside where they are made of a large mix of silicates and grow to sizes of 0.1 $\mu$m only.  Figure~\ref{fig:Teffpl1600K}  (4th panel) shows how the larger particles (red line) on the dayside fall faster (dotted line) through the atmosphere than the smaller cloud particles on the nightside.   Hence, cloud particle prevail for longer aloft on the nightside compared to the dayside. The size differences  will matter hugely for the albedo of the night- and the dayside of such a planet.

Cloud particles are a strong source of opacity in an atmosphere which will cause them to block our view into exoplanets and onto a potentially rocky surface. Cloud particles are also a chemically rather active component inside the gas that they are forming from. This is demonstrated in Figure~\ref{fig:Teffpl1600K} (5th and 6th row) in terms of the supersaturation ratio and the element abundances after the cloud particle have formed. 
 Section ~\ref{s:ccf} has touched the idea of the supersaturation of a gas phase which needs to be sufficiently high ($S\gg 1$)  in order to enable a phase transition, hence, the formation of cloud condensation nuclei. Figure~\ref{fig:Teffpl1600K} (5th) shows three regimes: $S\gg 1$,  $S=1$ and $S\ll 1$. $S\gg1$ occurs where all constructive cloud formation processes (nucleation and surface growth) take place. $S=1$ occurs where the growth process (which eventually supersedes the nucleation) can not progress any further and the respective materials are thermally stable. $S\ll 1$ indicates that here the materials that have condensed onto the cloud particles are thermally unstable and evaporate. Since cloud particles have grown to considerable size of 100$\mu$m on the nightside and 0.1 cm on the dayside at the cloud base, evaporation will not be instant but it occurs over a certain distance within the cloud base. At the cloud base,  a small increase of the element abundances (Fig.~\ref{fig:Teffpl1600K}, $\epsilon_x$, 4th) indicates that the evaporating cloud particles have transported elements from the upper into the lower atmospheric region. A considerable element depletion of the gas phase results from the cloud particle formation (seen as deep dips in $\epsilon_{\rm Ti}$, $\epsilon_{\rm Mg}$ etc.)  and a  decreased number density of all gas-phase species that form from those elements. Therefore, cloud particles do not only weaken the observability of spectroscopic features by adding a grey opacity source  for $\lambda < 1\mu$m, but they also weaken the spectral features by decreasing the number densities of species like TiO, VO, SiO and also H$_2$O.

\begin{figure} 
\centerline{\includegraphics[width=\textwidth]{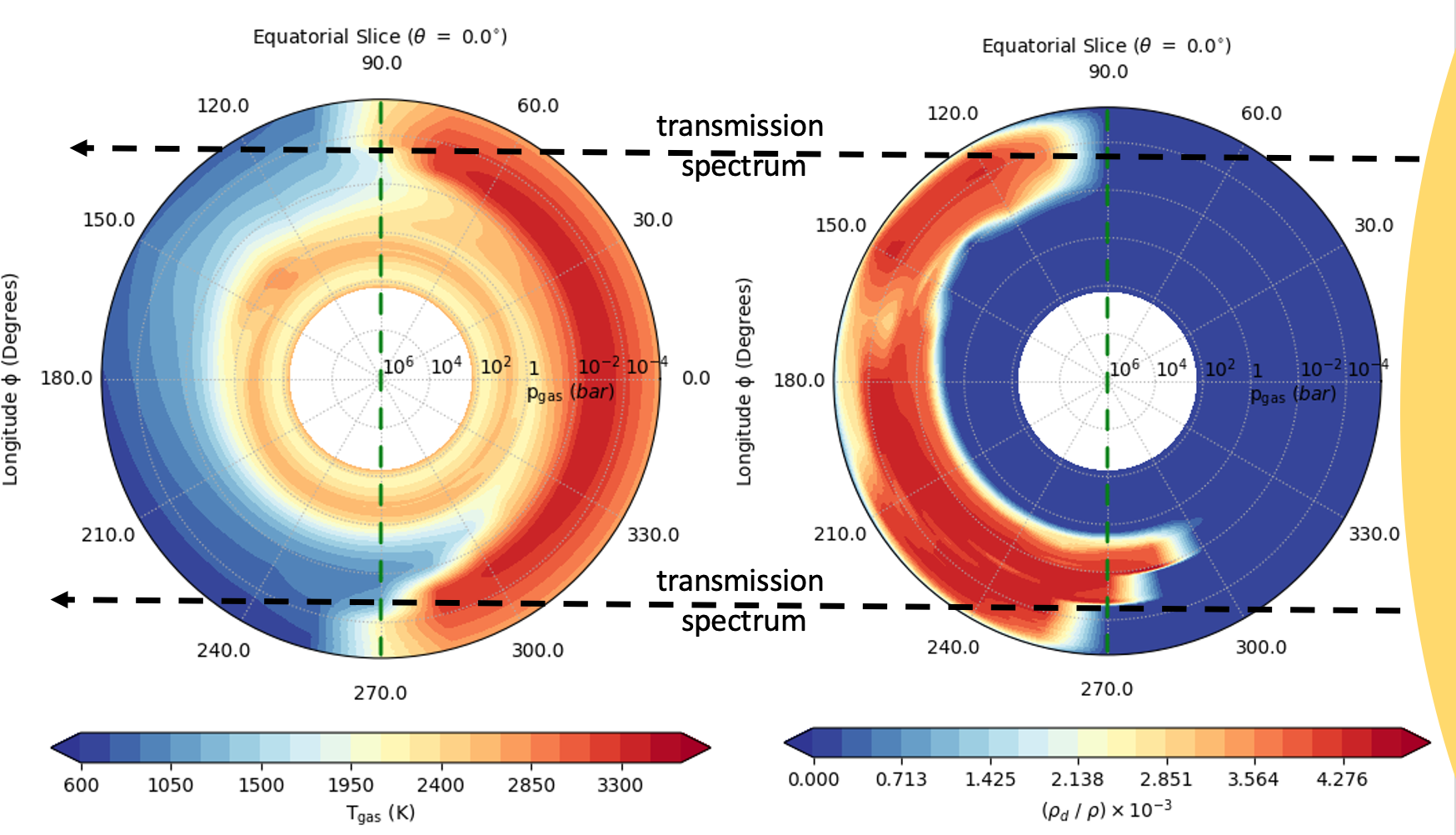}}
\vspace*{-0.5cm}
 \caption{Equatorial slices of a 3D planetary globe modelling the ultra-hot, tidally locked Jupiter HAT-7b. The location of the host star is indicated on the  right (yellow semicircle). The gas-phase temperature distribution is  on the left and the resulting  cloud mass load on the right. An extreme day/night asymmetry emerges for ultra-hot Jupiters which may be traced by transmission spectroscopy that measure the terminator regions. } 
 \label{fig:transmission}
 \end{figure}

\section{Moving forward}

We can now stich together a whole 3D cloud map by solving our cloud model for many such 1D profiles as in Fig.~\ref{fig:Teffpl1600K}. Clearly, we can not go into the amount of details discussed in Sect~\ref{s: results} but need to select key quantities. Figure~\ref{fig:transmission} shows two equatorial slices, for the gas temperature, $T_{\rm gas}$ [K],  which is the result of a cloud-free 3D GCM solution (left) and the cloud particle mass load (in terms of dust-to-gas mass ratios $\rho_d/\rho$) which is the result of our cloud formation simulation. Figure~\ref{fig:transmission} shows that clouds form only on the nightside on ultra-hot Jupiters and  that the amount of cloud particles is not homogeneous. The cloud reaches into the dayside on the morning terminator ($270^o$) , but not on the evening terminator ($90^o$).  The reason is an equatorial jet that transports cold gas from the nicht to the dayside. As cloud formation is determined by the local thermodynamic properties (unless in highly photon-dominated regions), our cloud map traces nicely this temperature difference and sharply drops where thermal instability disables the presence of clouds.  Note that  cloud formation is prohibited inside the blue regions on our cloud map (l.h.s. of Fig.~\ref{fig:transmission}).  Similarly, no clouds can from until the  hot dayside gas that is transported to the nightside has cooled sufficiently (e.g. like in Fig.~\ref{fig:Teffpl1600KSsat} )to allow cloud condensation seeds to form.  Figure~\ref{fig:transmission}  also suggests that if transmission spectra could distinguish between the morning and the evening terminator, a geometrical asymmetry should emerge such that the morning terminator appears more extended in the optical.

\section{Final thoughts}

Any astrophysical cloud formation model (e.g. for exoplanets and brown dwarfs) needs to be fundamental enough to enable the application to the wide variety of exoplanets that we already know about (rocky plants, ultra-hot gas giants, mini-Neptunes). For those, we have no solar-system analogous  to fly to and to take samples from. It is therefor essential that model approaches are compared \cite{2008MNRAS.391.1854H} and codes are not used a black boxes. Preferably, different methods are developed to solve the same problem\cite{2021arXiv210804701K}.  What is more, cloud formation requires the input from a variety of not-so adjacent research areas like quantum chemistry and geology, maybe even particle physics, and such fundamental research has been very time-consuming in the past. A summary of challenges was presented elsewhere\cite{2020arXiv201103302H}.

\medskip
\noindent
{\small {\bf Acknowledgment:} ChH thanks Dominic Samra and  David Lewis for their help with Figs.~\ref{fig:timescales},  \ref{fig:Teffpl1600KSsat} and  \ref{fig:Teffpl1600K}. ChH acknowledges funding from the European Union H2020-MSCA-ITN-2019 under Grant Agreement no. 860470 (CHAMELEON).}

\bibliographystyle{ws-rv-van}
\bibliography{references_clouds}

\begin{thebibliography}{40}
\providecommand{\natexlab}[1]{#1}
\providecommand{\url}[1]{\texttt{#1}}
\expandafter\ifx\csname urlstyle\endcsname\relax
  \providecommand{\doi}[1]{doi: #1}\else
  \providecommand{\doi}{doi: \begingroup \urlstyle{rm}\Url}\fi

\bibitem{2020MNRAS.497.5155W}
J.~{Wilson}, N.~P. {Gibson}, N.~{Nikolov}, S.~{Constantinou}, N.~{Madhusudhan},
  J.~{Goyal}, J.~K. {Barstow}, A.~L. {Carter}, E.~J.~W. {de Mooij},
  B.~{Drummond}, T.~{Mikal-Evans}, C.~{Helling}, N.~J. {Mayne}, and D.~K.
  {Sing}, {Ground-based transmission spectroscopy with FORS2: A featureless
  optical transmission spectrum and detection of H$_{2}$O for the ultra-hot
  Jupiter WASP-103b}, \emph{\mnras}. {\bf 497}\penalty0 (4), \penalty0
  5155--5170  (Oct., 2020).
\newblock \doi{10.1093/mnras/staa2307}.

\bibitem{2020AJ....160..280C}
K.~D. {Col{\'o}n}, L.~{Kreidberg}, L.~{Welbanks}, M.~R. {Line},
  N.~{Madhusudhan}, T.~{Beatty}, P.~{Tamburo}, K.~B. {Stevenson}, A.~{Mandell},
  J.~E. {Rodriguez}, T.~{Barclay}, E.~D. {Lopez}, K.~G. {Stassun},
  D.~{Angerhausen}, J.~J. {Fortney}, D.~J. {James}, J.~{Pepper}, J.~P.
  {Ahlers}, P.~{Plavchan}, S.~{Awiphan}, C.~{Kotnik}, K.~K. {McLeod},
  G.~{Murawski}, H.~{Chotani}, D.~{LeBrun}, W.~{Matzko}, D.~{Rea},
  M.~{Vidaurri}, S.~{Webster}, J.~K. {Williams}, L.~S. {Cox}, N.~{Tan}, and
  E.~A. {Gilbert}, {An Unusual Transmission Spectrum for the Sub-Saturn
  KELT-11b Suggestive of a Subsolar Water Abundance}, \emph{\aj}. 160\penalty0
  (6):\penalty0 280  (Dec., 2020).
\newblock \doi{10.3847/1538-3881/abc1e9}.

\bibitem{2021A&A...646A.168C}
L.~{Carone}, P.~{Molli{\`e}re}, Y.~{Zhou}, J.~{Bouwman}, F.~{Yan},
  R.~{Baeyens}, D.~{Apai}, N.~{Espinoza}, B.~V. {Rackham}, A.~{Jord{\'a}n},
  D.~{Angerhausen}, L.~{Decin}, M.~{Lendl}, O.~{Venot}, and T.~{Henning},
  {Indications for very high metallicity and absence of methane in the
  eccentric exo-Saturn WASP-117b}, \emph{\aap}. 646:\penalty0 A168  (Feb.,
  2021).
\newblock \doi{10.1051/0004-6361/202038620}.

\bibitem{2020A&A...639A...3C}
K.~L. {Chubb}, M.~{Min}, Y.~{Kawashima}, C.~{Helling}, and I.~{Waldmann},
  {Aluminium oxide in the atmosphere of hot Jupiter WASP-43b}, \emph{\aap}.
  639:\penalty0 A3  (July, 2020).
\newblock \doi{10.1051/0004-6361/201937267}.

\bibitem{2021A&A...646A..17B}
M.~{Braam}, F.~F.~S. {van der Tak}, K.~L. {Chubb}, and M.~{Min}, {Evidence for
  chromium hydride in the atmosphere of hot Jupiter WASP-31b}, \emph{\aap}.
  646:\penalty0 A17  (Feb., 2021).
\newblock \doi{10.1051/0004-6361/202039509}.

\bibitem{2019A&A...628A...9C}
N.~{Casasayas-Barris}, E.~{Pall{\'e}}, F.~{Yan}, G.~{Chen}, S.~{Kohl},
  M.~{Stangret}, H.~{Parviainen}, C.~{Helling}, N.~{Watanabe}, S.~{Czesla},
  A.~{Fukui}, P.~{Monta{\~n}{\'e}s-Rodr{\'\i}guez}, E.~{Nagel}, N.~{Narita},
  L.~{Nortmann}, G.~{Nowak}, J.~H.~M.~M. {Schmitt}, and M.~R. {Zapatero
  Osorio}, {Atmospheric characterization of the ultra-hot Jupiter
  MASCARA-2b/KELT-20b. Detection of CaII, FeII, NaI, and the Balmer series of H
  (H{\ensuremath{\alpha}}, H{\ensuremath{\beta}}, and H{\ensuremath{\gamma}})
  with high-dispersion transit spectroscopy}, \emph{\aap}. 628:\penalty0 A9
  (Aug., 2019).
\newblock \doi{10.1051/0004-6361/201935623}.

\bibitem{2020A&A...640C...6C}
N.~{Casasayas-Barris}, E.~{Pall{\'e}}, F.~{Yan}, G.~{Chen}, S.~{Kohl},
  M.~{Stangret}, H.~{Parviainen}, C.~{Helling}, N.~{Watanabe}, S.~{Czesla},
  A.~{Fukui}, P.~{Monta{\~n}{\'e}s-Rodr{\'\i}guez}, E.~{Nagel}, N.~{Narita},
  L.~{Nortmann}, G.~{Nowak}, J.~H.~M.~M. {Schmitt}, and M.~R. {Zapatero
  Osorio}, {Atmospheric characterization of the ultra-hot Jupiter
  MASCARA-2b/KELT-20b. Detection of Ca II, Fe II, Na I, and the Balmer series
  of H (H{\ensuremath{\alpha}}, H{\ensuremath{\beta}}, and
  H{\ensuremath{\gamma}}) with high-dispersion transit spectroscopy
  (Corrigendum)}, \emph{\aap}. 640:\penalty0 C6  (Aug., 2020).
\newblock \doi{10.1051/0004-6361/201935623e}.

\bibitem{2018NatAs...2..773S}
T.~S. {Stallard}, A.~G. {Burrell}, H.~{Melin}, L.~N. {Fletcher}, S.~{Miller},
  L.~{Moore}, J.~{O'Donoghue}, J.~E.~P. {Connerney}, T.~{Satoh}, and R.~E.
  {Johnson}, {Identification of Jupiter's magnetic equator through
  H$_{3}$$^{+}$ ionospheric emission}, \emph{Nature Astronomy}. {\bf 2},
  \penalty0 773--777  (July, 2018).
\newblock \doi{10.1038/s41550-018-0523-z}.

\bibitem{2020RvMP...92c5003M}
S.~{Miller}, J.~{Tennyson}, T.~R. {Geballe}, and T.~{Stallard}, {Thirty years
  of H$_{3}$$^{+}$ astronomy}, \emph{Reviews of Modern Physics}. 92\penalty0
  (3):\penalty0 035003  (July, 2020).
\newblock \doi{10.1103/RevModPhys.92.035003}.

\bibitem{2017pre8.conf..345H}
G.~{Hodosan}, C.~{Helling}, and P.~B. {Rimmer}.
\newblock {Exo-lightning radio emission: The case study of HAT-P-11b}.
\newblock In eds. G.~{Fischer}, G.~{Mann}, M.~{Panchenko}, and P.~{Zarka},
  \emph{Planetary Radio Emissions VIII}, pp. 345--356  (Jan., 2017).
\newblock \doi{10.1553/PRE8s345}.

\bibitem{2021MNRAS.502.6201B}
P.~{Barth}, C.~{Helling}, E.~E. {St{\"u}eken}, V.~{Bourrier}, N.~{Mayne}, P.~B.
  {Rimmer}, M.~{Jardine}, A.~A. {Vidotto}, P.~J. {Wheatley}, and R.~{Fares},
  {MOVES - IV. Modelling the influence of stellar XUV-flux, cosmic rays, and
  stellar energetic particles on the atmospheric composition of the hot Jupiter
  HD 189733b}, \emph{\mnras}. {\bf 502}\penalty0 (4), \penalty0 6201--6215
  (Apr., 2021).
\newblock \doi{10.1093/mnras/staa3989}.

\bibitem{2016MNRAS.461.3927H}
G.~{Hodos{\'a}n}, C.~{Helling}, R.~{Asensio-Torres}, I.~{Vorgul}, and P.~B.
  {Rimmer}, {Lightning climatology of exoplanets and brown dwarfs guided by
  Solar system data}, \emph{\mnras}. {\bf 461}\penalty0 (4), \penalty0
  3927--3947  (Oct., 2016).
\newblock \doi{10.1093/mnras/stw1571}.

\bibitem{2018AJ....155...29W}
H.~R. {Wakeford}, D.~K. {Sing}, D.~{Deming}, N.~K. {Lewis}, J.~{Goyal}, T.~J.
  {Wilson}, J.~{Barstow}, T.~{Kataria}, B.~{Drummond}, T.~M. {Evans}, A.~L.
  {Carter}, N.~{Nikolov}, H.~A. {Knutson}, G.~E. {Ballester}, and A.~M.
  {Mandell}, {The Complete Transmission Spectrum of WASP-39b with a Precise
  Water Constraint}, \emph{\aj}. 155\penalty0 (1):\penalty0 29  (Jan., 2018).
\newblock \doi{10.3847/1538-3881/aa9e4e}.

\bibitem{2020MNRAS.497.4183B}
J.~K. {Barstow}, {Unveiling cloudy exoplanets: the influence of cloud model
  choices on retrieval solutions}, \emph{\mnras}. {\bf 497}\penalty0 (4),
  \penalty0 4183--4195  (Oct., 2020).
\newblock \doi{10.1093/mnras/staa2219}.

\bibitem{2021ApJ...913..114W}
L.~{Welbanks} and N.~{Madhusudhan}, {Aurora: A Generalized Retrieval Framework
  for Exoplanetary Transmission Spectra}, \emph{\apj}. 913\penalty0
  (2):\penalty0 114  (June, 2021).
\newblock \doi{10.3847/1538-4357/abee94}.

\bibitem{2016IJAsB..15...45B}
S.~V. {Berdyugina}, J.~R. {Kuhn}, D.~M. {Harrington}, T.~{{\v{S}}antl-Temkiv},
  and E.~J. {Messersmith}, {Remote sensing of life: polarimetric signatures of
  photosynthetic pigments as sensitive biomarkers}, \emph{International Journal
  of Astrobiology}. {\bf 15}\penalty0 (1), \penalty0 45--56  (Jan., 2016).
\newblock \doi{10.1017/S1473550415000129}.

\bibitem{2014IJAsB..13..165S}
C.~R. {Stark}, C.~{Helling}, D.~A. {Diver}, and P.~B. {Rimmer}, {Electrostatic
  activation of prebiotic chemistry in substellar atmospheres},
  \emph{International Journal of Astrobiology}. {\bf 13}\penalty0 (2),
  \penalty0 165--172  (Apr, 2014).
\newblock \doi{10.1017/S1473550413000475}.

\bibitem{2021Univ....7..172S}
S.~{Seager}, J.~J. {Petkowski}, M.~N. {G{\"u}nther}, W.~{Bains},
  T.~{Mikal-Evans}, and D.~{Deming}, {Possibilities for an Aerial Biosphere in
  Temperate Sub Neptune-Sized Exoplanet Atmospheres}, \emph{Universe}. {\bf
  7}\penalty0 (6), \penalty0 172  (May, 2021).
\newblock \doi{10.3390/universe7060172}.

\bibitem{2019AREPS..47..583H}
C.~{Helling}, {Exoplanet Clouds}, \emph{Annual Review of Earth and Planetary
  Sciences}. {\bf 47}, \penalty0 583--606  (May, 2019).
\newblock \doi{10.1146/annurev-earth-053018-060401}.

\bibitem{2021JGRE..12606655G}
P.~{Gao}, H.~R. {Wakeford}, S.~E. {Moran}, and V.~{Parmentier}, {Aerosols in
  Exoplanet Atmospheres}, \emph{Journal of Geophysical Research (Planets)}.
  126\penalty0 (4):\penalty0 e06655  (Apr., 2021).
\newblock \doi{10.1029/2020JE006655}.

\bibitem{2021MNRAS.502.2198T}
X.~{Tan} and A.~P. {Showman}, {Atmospheric circulation of brown dwarfs and
  directly imaged exoplanets driven by cloud radiative feedback: global and
  equatorial dynamics}, \emph{\mnras}. {\bf 502}\penalty0 (2), \penalty0
  2198--2219  (Apr., 2021).
\newblock \doi{10.1093/mnras/stab097}.

\bibitem{2010A&A...513A..56G}
C.~{G{\"u}ttler}, J.~{Blum}, A.~{Zsom}, C.~W. {Ormel}, and C.~P. {Dullemond},
  {The outcome of protoplanetary dust growth: pebbles, boulders, or
  planetesimals?. I. Mapping the zoo of laboratory collision experiments},
  \emph{\aap}. 513:\penalty0 A56  (Apr, 2010).
\newblock \doi{10.1051/0004-6361/200912852}.

\bibitem{2016JGRA..121.8152S}
J.~{Svensmark}, M.~B. {Enghoff}, N.~J. {Shaviv}, and H.~{Svensmark}, {The
  response of clouds and aerosols to cosmic ray decreases}, \emph{Journal of
  Geophysical Research (Space Physics)}. {\bf 121}\penalty0 (9), \penalty0
  8152--8181  (Sept., 2016).
\newblock \doi{10.1002/2016JA022689}.

\bibitem{2019JPhCS1322a2028H}
C.~{Helling}.
\newblock {Lightning in other planets}.
\newblock In \emph{Journal of Physics Conference Series}, vol. 1322,
  \emph{Journal of Physics Conference Series}, p. 012028  (Oct., 2019).
\newblock \doi{10.1088/1742-6596/1322/1/012028}.

\bibitem{2009A&A...506.1367W}
S.~{Witte}, C.~{Helling}, and P.~H. {Hauschildt}, {Dust in brown dwarfs and
  extra-solar planets. II. Cloud formation for cosmologically evolving
  abundances}, \emph{\aap}. {\bf 506}\penalty0 (3), \penalty0 1367--1380  (Nov,
  2009).
\newblock \doi{10.1051/0004-6361/200811501}.

\bibitem{2003A&A...399..297W}
P.~{Woitke} and C.~{Helling}, {Dust in brown dwarfs. II. The coupled problem of
  dust formation and sedimentation}, \emph{\aap}. {\bf 399}, \penalty0 297--313
   (Feb., 2003).
\newblock \doi{10.1051/0004-6361:20021734}.

\bibitem{2004A&A...423..657H}
C.~{Helling}, R.~{Klein}, P.~{Woitke}, U.~{Nowak}, and E.~{Sedlmayr}, {Dust in
  brown dwarfs. IV. Dust formation and driven turbulence on mesoscopic scales},
  \emph{\aap}. {\bf 423}, \penalty0 657--675  (Aug., 2004).
\newblock \doi{10.1051/0004-6361:20034514}.

\bibitem{2013RSPTA.37110581H}
C.~{Helling} and A.~{Fomins}, {Modelling the formation of atmospheric dust in
  brown dwarfs and planetary atmospheres}, \emph{Philosophical Transactions of
  the Royal Society of London Series A}. {\bf 371}\penalty0 (1994), \penalty0
  20110581--20110581  (June, 2013).
\newblock \doi{10.1098/rsta.2011.0581}.

\bibitem{2013pccd.book.....G}
H.-P. {Gail} and E.~{Sedlmayr}, \emph{{Physics and Chemistry of Circumstellar
  Dust Shells}}  (2013).

\bibitem{1998A&A...337..847P}
A.~B.~C. {Patzer}, A.~{Gauger}, and E.~{Sedlmayr}, {Dust formation in stellar
  winds. VII. Kinetic nucleation theory for chemical non-equilibrium in the gas
  phase}, \emph{\aap}. {\bf 337}, \penalty0 847--858  (Sep, 1998).

\bibitem{2021arXiv210804701K}
C.~{K{\"o}hn}, C.~{Helling}, M.~{B{\o}dker Enghoff}, K.~{Haynes}, D.~{Krog},
  J.~P. {Sindel}, and D.~{Gobrecht}, {Dust in brown dwarfs and extra-solar
  planets. VIII. TiO$_2$ seed formation: 3D Monte Carlo versus kinetic
  approach}, \emph{arXiv e-prints}. art. arXiv:2108.04701  (Aug., 2021).

\bibitem{1996ASPC...96...69G}
A.~{Goeres}.
\newblock {Chemistry and thermodynamics of the nucleation in R CrB star
  shells}.
\newblock In eds. C.~S. {Jeffery} and U.~{Heber}, \emph{Hydrogen Deficient
  Stars}, vol.~96, \emph{Astronomical Society of the Pacific Conference
  Series}, p.~69  (Jan., 1996).

\bibitem{2018A&A...614A.126L}
E.~K.~H. {Lee}, J.~{Blecic}, and C.~{Helling}, {Dust in brown dwarfs and
  extra-solar planets. VI. Assessing seed formation across the brown dwarf and
  exoplanet regimes}, \emph{\aap}. 614:\penalty0 A126  (Jul, 2018).
\newblock \doi{10.1051/0004-6361/201731977}.

\bibitem{2018ApJ...860...18P}
D.~{Powell}, X.~{Zhang}, P.~{Gao}, and V.~{Parmentier}, {Formation of Silicate
  and Titanium Clouds on Hot Jupiters}, \emph{\apj}. 860\penalty0 (1):\penalty0
  18  (June, 2018).
\newblock \doi{10.3847/1538-4357/aac215}.

\bibitem{2019ApJ...876L...5K}
Y.~{Kawashima}, R.~{Hu}, and M.~{Ikoma}, {Detectable Molecular Features above
  Hydrocarbon Haze via Transmission Spectroscopy with JWST: Case Studies of GJ
  1214b-, GJ 436b-, HD 97658b-, and Kepler-51b-like Planets}, \emph{\apjl}.
  876\penalty0 (1):\penalty0 L5  (May, 2019).
\newblock \doi{10.3847/2041-8213/ab16f6}.

\bibitem{2020A&A...641A.178H}
C.~{Helling}, Y.~{Kawashima}, V.~{Graham}, D.~{Samra}, K.~L. {Chubb}, M.~{Min},
  L.~B.~F.~M. {Waters}, and V.~{Parmentier}, {Mineral cloud and hydrocarbon
  haze particles in the atmosphere of the hot Jupiter JWST target WASP-43b},
  \emph{\aap}. 641:\penalty0 A178  (Sept., 2020).
\newblock \doi{10.1051/0004-6361/202037633}.

\bibitem{2021MNRAS.tmp.1277B}
R.~{Baeyens}, L.~{Decin}, L.~{Carone}, O.~{Venot}, M.~{Ag{\'u}ndez}, and
  P.~{Molli{\'e}re}, {Grid of Pseudo-2D chemistry models for tidally-locked
  exoplanets. I. The role of vertical and horizontal mixing}, \emph{\mnras}
  (May, 2021).
\newblock \doi{10.1093/mnras/stab1310}.

\bibitem{2021A&A...649A..44H}
C.~{Helling}, D.~{Lewis}, D.~{Samra}, L.~{Carone}, V.~{Graham}, O.~{Herbort},
  K.~L. {Chubb}, M.~{Min}, R.~{Waters}, V.~{Parmentier}, and N.~{Mayne}, {Cloud
  property trends in hot and ultra-hot giant gas planets (WASP-43b, WASP-103b,
  WASP-121b, HAT-P-7b, and WASP-18b)}, \emph{\aap}. 649:\penalty0 A44  (May,
  2021).
\newblock \doi{10.1051/0004-6361/202039911}.

\bibitem{2008MNRAS.391.1854H}
C.~{Helling}, A.~{Ackerman}, F.~{Allard}, M.~{Dehn}, P.~{Hauschildt},
  D.~{Homeier}, K.~{Lodders}, M.~{Marley}, F.~{Rietmeijer}, T.~{Tsuji}, and
  P.~{Woitke}, {A comparison of chemistry and dust cloud formation in ultracool
  dwarf model atmospheres}, \emph{\mnras}. {\bf 391}\penalty0 (4), \penalty0
  1854--1873  (Dec., 2008).
\newblock \doi{10.1111/j.1365-2966.2008.13991.x}.

\bibitem{2020arXiv201103302H}
C.~{Helling}, {Clouds in Exoplanetary Atmospheres}, \emph{arXiv e-prints}. art.
  arXiv:2011.03302  (Nov., 2020).

\end{thebibliography}
\end{document}